\documentclass[aps,prb,twocolumn,groupedaddress,amssymb,showpacs]{revtex4}
\usepackage{graphicx}
\usepackage{bm}

\begin{document}

\title{Surface impedance anisotropy of YBa$_2$Cu$_3$O$_{6.95}$ single crystals:
electrodynamic basis of the measurements}

\author{Yu.A.~Nefyodov} \email{nefyodov@issp.ac.ru}
\author{M.R.~Trunin}
\author{A.A.~Zhohov}
\author{I.G.~Naumenko}
\author{G.A.~Emel'chenko}
\affiliation{Institute of
Solid State Physics RAS, 142432 Chernogolovka, Moscow district,
Russia}
\author{D.Yu.~Vodolazov}
\author{I.L.~Maksimov}
\affiliation{Nizhny Novgorod University, 23 Gagarin Avenue,
Nizhny Novgorod 603600, Russia}

%\date{submitted to Phys.Rev.B }

\begin{abstract}
An electrodynamic technique is developed for determining the components of
surface impedance and complex conductivity tensors of HTSC single crystals
on the basis of measured quantities of a quality factor and a resonator
frequency shift. A simple formula is obtained for a geometrical factor of
a crystal in the form of a plate with dimensions $b\gg a>c$ in a microwave
magnetic field ${\bf H_{\omega}}\perp ab$. To obtain the $c$-axis complex
conductivity from measurements at ${\bf H_{\omega}}\parallel ab$ we
propose a procedure which takes account of sample size effects. With the
aid of the technique involved temperature dependences of all impedance and
conductivity tensors components of YBa$_2$Cu$_3$O$_{6.95}$ single crystal,
grown in BaZrO$_3$ crucible, are determined at a frequency of $f=9.4$~GHz
in its normal and superconducting states. All of them proved to be linear
at $T<T_c/2$, and their extrapolation to zero temperature gives the values
of residual surface resistance $R_{ab}(0)\approx 40$~$\mu\Omega$ and
$R_c(0)\approx 0.8$~m$\Omega$ and magnetic field penetration depth
$\lambda_{ab}(0)\approx 150$~nm and $\lambda_c(0)\approx 1.55$~$\mu$m.
\end{abstract}

% 74.72.Bk Y-based cuprates
 \maketitle

\section{Introduction}
Microwave measurements of the temperature dependence of the
complex conductivity tensor
$\hat\sigma(T)=\hat\sigma'(T)-i\hat\sigma''(T)$ of
high-$T_c$ superconductors (HTSC) have advanced
considerably our understanding of the mechanisms of
quasiparticles transport along crystallographic axes of
these anisotropic compounds. The real part $\hat\sigma'(T)$
is susceptible to the scattering rate of quasiparticles, as
well as their density of states. The imaginary part
$\hat\sigma''(T)$ is related to the magnetic field
penetration depth $\lambda(T)$. In the local
electrodynamics, which can be applied to HTSC,
\begin{equation}
\hat{\sigma}(T)=i\omega\mu_0/\hat{Z^2}(T),
\end{equation}
where $\hat{Z}(T)$ is the surface impedance tensor of the
sample, $\omega=2\pi f$ and $\mu_0=4\pi\cdot10^{-7}$~H/m.
In HTSC the tensors $\hat{Z}$ and $\hat{\sigma}$ are
characterized by two components: $Z_{ab} =R_{ab}+iX_{ab}$
(or $\sigma_{ab}=\sigma'_{ab}-i\sigma''_{ab}$) in weakly
anisotropic $ab$-planes CuO$_2$ and $Z_c=R_c+iX_c$
($\sigma_c=\sigma'_c-i\sigma''_c$) perpendicular to these
planes.

In the temperature range $T\ge T_c$, $R_{ab}(T)=X_{ab}(T)$ in the
$ab$-plane of the optimum-doped YBa$_2$Cu$_3$O$_{6.95}$
\cite{Shi,Ach,Kit1,Tru6,Shi1} and
Bi$_2$Sr$_2$CaCu$_2$O$_{8+\delta}$ \cite{Shi1,Jac1,Shov} single
crystals, and this relation is equivalent to the condition of the
normal skin effect. The common features of these crystals are the
linear temperature dependence of the surface resistance ($\Delta
R_{ab}(T)\propto T$) and of the surface reactance ($\Delta
X_{ab}(T)\propto \Delta \lambda_{ab}(T)\propto T$) at temperatures
$T\ll T_c$ (see Refs.~\onlinecite{Bonn4,Tru2,Trunef,Trugol} and
references therein). The difference is that the linear resistivity
region extends to near $T_c/2$ for
Bi$_2$Sr$_2$CaCu$_2$O$_{8+\delta}$ and terminates near or below
$T<T_c/3$ for YBa$_2$Cu$_3$O$_{6.95}$ single crystals. At higher
temperatures $R_s(T)$ of YBa$_2$Cu$_3$O$_{6.95}$ has a broad peak.
In addition, the $\lambda _{ab}(T)$ curves of some
YBa$_2$Cu$_3$O$_{6.95}$ single crystals have unusual features in
the intermediate temperature range \cite{Tru6,Srik1}.

In comparison with the microwave response of the cuprate layers of HTSC,
the data concerning their microwave properties in the direction
perpendicular to these layers are scarce. Moreover, the available
experimental data are controversial. In this connection, the major
electrodynamic problem is the accuracy of the techniques used in
determination of $Z_c(T)$ and $\sigma_c(T)$ in HTSC.

The most convenient technique for measurements of the surface impedance of
small HTSC samples in the X-W microwave frequency bands is the so-called
`hot-finger' method \cite{Tru2,Sri7}. The underlying idea of the method is
that a crystal is set on a sapphire rod at the center of a superconducting
cylindrical cavity resonating at the frequency $f$ in the $H_{011}$ mode,
i.e., at the antinode of a quasi-homogeneous microwave magnetic field. In
the experiment the real $R$ and imaginary $X$ parts of the surface
impedance are derived from the following relations:
\begin{equation}
R=\Gamma\,\Delta(1/Q),\qquad X=-2\,\Gamma\,\delta f/f. \label{RX}
\end{equation}

Here $\Gamma$ is the sample geometrical factor; $\Delta (1/Q)$ is the
difference between the values 1/Q of the cavity with the sample inside and
the empty one; $\delta f$ is the frequency shift relative to that which
would be measured for a sample with perfect screening and no penetration
of microwave fields. In the experiment we measure the difference $\Delta
f(T)$ between resonant frequency shifts versus temperature of the loaded
and empty cavity, which is equal to $\Delta f(T)=\delta f(T)+f_0$, where
$f_0$ is a constant. In HTSC single crystals the constant $f_0$ can be
determined from measurements of the surface impedance in the normal state
\cite{Tru1}. Another quantity essential for determining the values of
$R(T)$ and $X(T)$ from Eq.~(2) is the sample geometric factor
\begin{equation}
\Gamma=\frac{2\omega W}{\gamma}\,, \quad
W=\frac{\mu_0}{2}\int_V H_{\omega}^2dV, \quad \gamma=\int_s
H_t^2ds, \label{gamma}
\end{equation}
where $W$ is the energy stored in the cavity, $V$ is the
volume of the cavity, $\bf{H_{\omega}}$ is the microwave
magnetic field generated in the cavity, $s$ is the total
sample surface area, and ${\bf H}_t$ is the tangential
component of $\bf{H_{\omega}}$ on the sample surface. The
energy $W$ is easily obtained for the resonator mode under
use, therefore the task of deriving the impedance value is
reduced to defining the integral $\gamma$ in Eq.~(3). The
task simplifies if a typical HTSC crystal in the form of a
rectangular plate with dimensions $b\gg a>c$ and volume
$v\sim 0.1$~mm$^3$ is radiated by the microwave magnetic
field ${\bf H_{\omega}}\parallel b$ ($L$-orientation,
Fig.~1a).
%fig.1-----------------------------------------
\begin{figure}[hbtp]
\includegraphics[width=0.4\textwidth,clip]{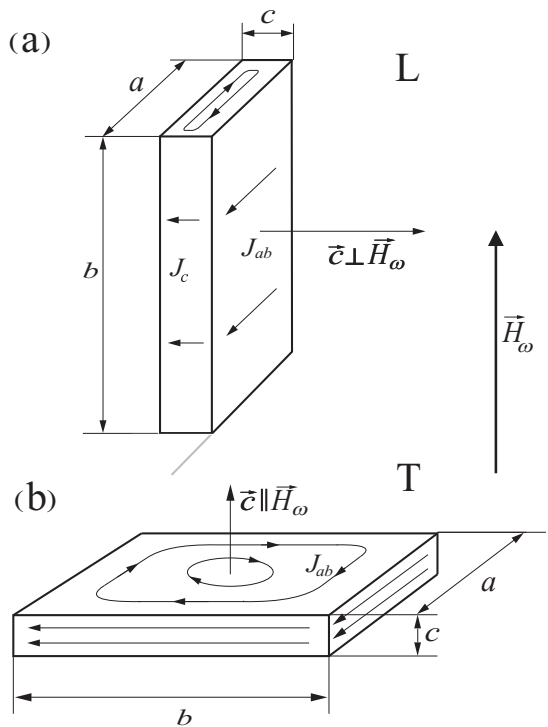}
%\vspace{0.2cm}
\caption{Two experimental orientations of the sample with respect
to the microwave magnetic field ${\bf H}_\omega$: a) longitudinal
$L$-orientation, ${\bf H}_\omega\parallel ab$, and b) transverse
$T$-orientation, ${\bf H}_\omega\perp ab$. Arrows indicate the
direction of
the high-frequency currents.} \label{f1} %\vspace{0cm}
\end{figure}
If this is the case, in the superconducting state at $T<0.9\,T_c$,
when the magnetic field penetration depth is smaller than the
characteristic sample dimensions, the quantity $H_t\simeq H_0$ can
be taken out of the integral $\gamma$, where $H_0$ is an amplitude
of ${\bf H_{\omega}}$. Thus we obtain $\gamma\simeq
2H_0^2(ab+bc)$, so the surface impedance $Z_{ab+c}$ for the sample
in the $L$-orientation will be equal to
\begin{equation}
Z_{ab+c}\simeq \frac{abZ_{ab}+bcZ_c}{ab+bc}, \label{Zabc}
\end{equation}
where the subscripts of $Z$ denote the directions of the
screening currents. While deriving formula (4) we neglect
not only weak anisotropy in the $ab$-plane but also a
contribution of the crystal $ac$-faces which is apparently
minute in comparison with the summands in the numerator in
Eq.~(4) due to the areas difference $ac\ll bc<ab$. The
cleaving of the crystal along the $b$-edge into several
needles multiplies up the contribution from the $c$-axis,
so that measurements of $Z_{ab+c}$ before and after the
cleaving allowed to extract $Z_c$ in the superconducting
state of YBa$_2$Cu$_3$O$_{6.95}$ single crystal \cite{Hos}.
However, this procedure has the following disadvantages:
(i) it assumes $abZ_{ab}$ term in Eq.~(4) to be
non-alterable, which results in an uncontrolled inaccuracy
due to non-ideal sample cleaving into rectangular needles,
(ii) as it will be shown below, the size effect takes place
at temperatures $T>0.9\,T_c$ in the $L$-orientation; this
restricts applicability range of Eq.~(4) within low
temperatures and does not allow to extract $\lambda_c(0)$
value, and (iii) in many cases one needs to save the
initial sample for further study, e.g., for investigation
of evolution of its anisotropic properties with doping
level.

Therefore, consecutive measurements of the crystal at first in
transverse ($T$) orientation ${\bf H_{\omega}}\parallel c$
(Fig.~1b) to obtain $Z_{ab}$ and then in longitudinal ($L$) one
appear to be a more natural way to obtain $Z_c$ value. A
difficulty in determining of the geometrical factor $\Gamma$ or
the integral $\gamma$ in Eq.~(3) in the $T$-orientation of the
crystal arises while using this technique. As mentioned in
Ref.~\onlinecite{Tru2}, $\gamma\simeq 2H_0^2a^2[\ln (a/c)+1]$
proves to be a reasonable estimation for a square sample with
$a=b\gg c$. It is also known that the approximation of a
rectangular plate to an ellipsoid inscribed in it results in an
overestimated value of $\gamma$.

The purpose of this paper is (i) to calculate $\gamma$ for
a typical HTSC crystal in the $T$-orientation, (ii) to
generalize formula (4) for the $L$-orientation of the
crystal to the range of higher temperatures $T>0.9\,T_c$,
and (iii) to report on the measurement results for all
surface impedance components of high-quality
YBa$_2$Cu$_3$O$_{6.95}$ single crystals, grown in BaZrO$_3$
crucibles, in the normal and superconducting states.

\section{Geometrical factor in the $T$-orientation}

Let us consider a rectangular ideal conductor with dimensions
$L_y\gg L_x, L_z$ placed in a constant magnetic field ${\bf
H}\parallel z$ (Fig.~2). The problem of obtaining the field
distribution around such a conductor becomes two-dimensional and
to solve it one can apply the method suggested in
Ref.~\onlinecite{Provost}. The magnetic field outside the
conductor satisfies Maxwell equations ${\bf\nabla}\times{\bf H}=0$
and ${\bf\nabla\bf B}=0$. The former equation allows to introduce
a scalar potential $\varphi$, and the latter allows to introduce a
vector potential ${\bf A}$: ${\bf
H}=-{\bf\nabla\varphi}={\bf\nabla}\times{\bf A}/\mu_0$. Let ${\bf
A}$ be directed along the $y$-axis: ${\bf A}=(0,A,0)$.
%fig.2-----------------------------------------
\begin{figure}[tp]
\includegraphics[width=0.44\textwidth,clip]{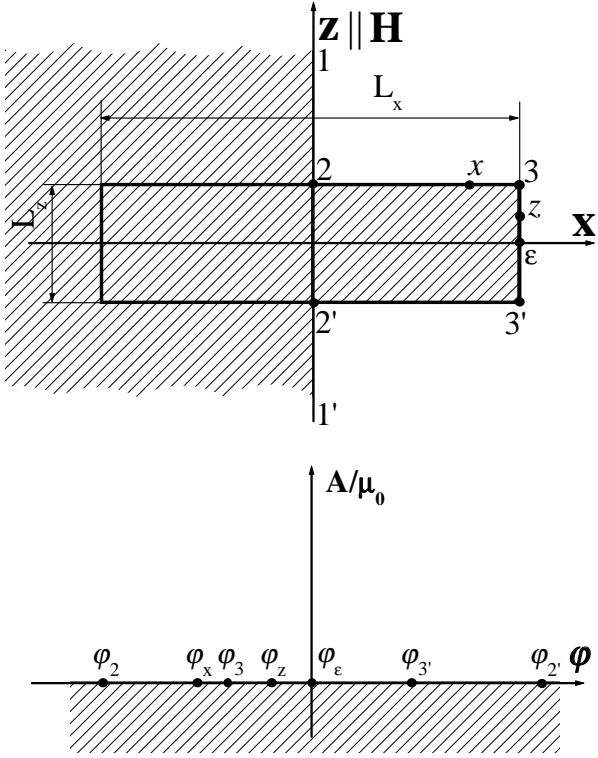}
%\vspace{0cm}
\caption{Complex planes $(x,z)$ and $(\varphi,A/\mu_0)$ used for
conformal mapping.} \label{f2}
\end{figure}
Then the magnetic field components will be as follows \cite{Lan}:
\begin{equation}
H_x-iH_z=-\frac{d\psi}{dw}, \label{h}
\end{equation}
where the complex potential $\psi=\varphi+iA/\mu_0$ is an
analytical function of $w=x+iz$ variable, which determines
the conformal mapping of $(x,z)$ plane into
$(\varphi,A/\mu_0)$ plane. Schwarz transformation
\begin{equation}
\frac{dw}{d\psi}=-\frac{i}{H}\sqrt{\frac{(\psi-\psi_3)(\psi-\psi_{3'})}
{(\psi-\psi_2)(\psi-\psi_{2'})}} \label{swarz}
\end{equation}
specifies the relationship between the unshaded areas in
Fig.~2. Placing the point $\psi_\epsilon$ into plain
$(\varphi,A/\mu_0)$ point of origin, we get
$\varphi_{\epsilon}=0$ and $A=0$ ($\psi=\varphi$) along the
path
$1\rightarrow2\rightarrow3\rightarrow\epsilon\rightarrow3'\rightarrow2'\rightarrow1'$.
Owing to the task symmetry $\varphi_2=-\varphi_{2'}$ and
$\varphi_3=-\varphi_{3'}$. Integrating Eq.~(\ref{swarz}),
we obtain the following relationship between $x,z$
coordinates and potential $\varphi$ on the conductor
surface:
\begin{eqnarray}
x=-\frac{\varphi_3}{H}\frac{\sqrt{1-k^2}}{k}E\left(\sqrt{1-\frac{\varphi^2}{\varphi_2^2}},
\frac{1}{\sqrt{1-k^2}}\right),\nonumber\\(z=L_z/2, 0\leq
x\leq L_x/2),\nonumber \\
z=-\frac{\varphi_3}{H}E\left(\frac{\varphi}{\varphi_2},\frac{1}{k}\right),\quad(x=L_x/2,
0\leq z\leq L_z/2), \label{xz}
%\nonumber
\end{eqnarray}
where $k=\varphi_3/\varphi_2$, and $E(u,v)$ is an incomplete elliptic
integral of the second kind. Eqs.~(\ref{h}) and (\ref{xz}) define an
implicit  dependence of the magnetic field against coordinates on the
conductor surface. The upper inset in Fig.~3 displays the distribution of
$H_z(z)$ and $H_x(x)$ magnetic field components in $L_zL_y$ and $L_xL_y$
planes, respectively, for three different $L_z/L_x$ ratios.

From the formulae (5)-(7) one can easily calculate the
magnetic moment of the conductor
\begin{eqnarray}
M=\left |\frac{1}{2}\int{{\bf j}\times {\bf r}\,dv}\right |
=\nonumber\\
=-4L_y\left(\int_0^{L_x/2}x|H_x|dx+\frac{L_x}{2}\int_{0}^{L_z/2}H_zdz\right)
\label{m}
\end{eqnarray}
and the integral $\gamma$ in Eq.~(3):
\begin{equation}
\gamma=4L_y\left(\int_0^{L_x/2}H_x^2dx+\int_0^{L_z/2}H_z^2dz\right).\label{g}
\end{equation}

In order to obtain this we change from integration over
coordinates in Eqs.~(8) and (9) to integration over
potential, taking formulae (5) and (6) into account:
\begin{eqnarray}
-\frac{M}{L_yH}=\frac{4}{H}\left(\int_{\varphi_2}^{\varphi_3}xd\varphi-
\frac{L_x}{2}\int_0^{\varphi_3}d\varphi\right)=\nonumber\\=
\frac{\pi(1-k^2)L_z^2}{4}\left(E(k)-(1-k^2)K(k)\right)^2,
\label{mom}
\end{eqnarray}

\begin{eqnarray}
\frac{\gamma}{L_yH^2}=\frac{4}{H}\left(
\int_{\varphi_2}^{\varphi_3}\sqrt{\frac{\varphi_2^2-\varphi^2}{\varphi^2-\varphi_3^2}}d\varphi+
\right.&& \nonumber \\\left.+
\int_{\varphi_3}^{0}\sqrt{\frac{\varphi_2^2-\varphi^2}{\varphi_3^2-\varphi^2}}d\varphi\right)=
2\left(L_xf_x+L_zf_z\right),&&
\end{eqnarray}
where
\begin{eqnarray}
f_x=\frac{K(\sqrt{1-k^2})-E(\sqrt{1-k^2})}{E(\sqrt{1-k^2})-k^2K(\sqrt{1-k^2})},&&\nonumber
\\ f_z= \frac{E(k)}{E(k)-(1-k^2)K(k)},&&
\end{eqnarray}
with $K(v)$ and $E(v)$ being complete elliptic integrals of the
first and second kinds. Formula (10) coincides with the result
obtained previously in Ref.~\onlinecite{Brandt}. Furthermore, the
relationship between the ratio $L_z/L_x$ and $k$ is required to
calculate the values of the moment $M/v$ ($v=L_xL_yL_z$ is a
conductor volume) and the factor $\gamma/s$ ($s=2L_y(L_z+L_x)$ is
a conductor surface area), which we obtain substituting the values
$x=L_x/2$ and $z=L_z/2$ into the left side of Eq.~(7):
\begin{equation}
\frac{L_z}{L_x}=\frac{E(k)-(1-k^2)K(k)}{E(\sqrt{1-k^2})-k^2K(\sqrt{1-k^2})}.
\label{f2f3}
\end{equation}

%fig.3-----------------------------------------
\begin{figure}[tp]
\includegraphics[width=0.45\textwidth,clip]{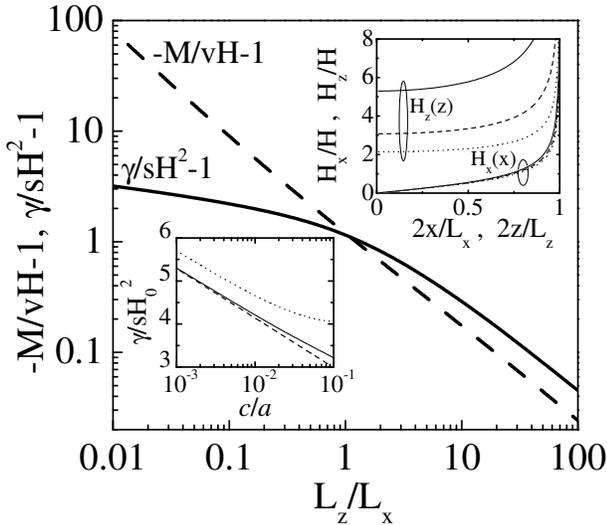}
%\vspace{0.2cm}
\caption{The values of $[-M/vH -1]$ (dashed line) and
$[\gamma/sH^2-1]$ (solid line) calculated from Eqs.~(10) and (11)
versus the ratio $L_z/L_x$. The upper inset represents the
magnetic field distribution on the surfaces $L_zL_y$ (upper three
curves) and $L_xL_y$ (lower curves) for the values of
$L_z/L_x=0.25$ (dotted lines), 0.1 (dashed lines), and 0.03 (solid
lines). The lower inset: $\gamma/sH_0^2$ as a function of $c/a$.
Solid and dashed lines correspond to the calculations from
Eq.~(11) and Eq.~(14), respectively. Dotted line is the upper
limit of $\gamma/sH_0^2$ estimated from Eq.~(15) for $b=4a$.}
\label{f3}
\end{figure}

$M$ and $\gamma$ dependences on $L_z/L_x$ ratio, computed
from formulae (10) and (11), are presented in Fig.~3. If
$L_z\gg L_x$, the magnitude of the magnetic field on the
sample surface tends to that of the applied one along with
the values of $M$ and $\gamma$ tending to $(-vH)$ and
$sH^2$, respectively.

Presently we apply the formulae obtained to determining the
HTSC crystal geometrical factor when transversely placed
with respect to ${\bf H_{\omega}}$ field (Fig.~1b). The
field distribution on the ideal conductor surface will
coincide with alternating magnetic field distribution on
the surface of a superconducting sample of the same
dimensions ($L_x=a$, $L_y=b$, $L_z=c$) placed in microwave
field ${\bf H_\omega}={\bf H}$, provided that the
penetration depth is smaller than the sample dimensions. It
is easy to obtain a simple estimation of $\gamma$ in case
of a very thin crystal ($c\ll a\ll b$). Upon keeping the
first term from the expansion into $k\ll 1$ series of the
right side of Eq.~(13), we get the value
$k\simeq\sqrt{4L_z/\pi L_x}$. Therefore, small values of
$k$ correspond to $c/a\ll 1$. The subsequent substitution
of $k=\sqrt{4c/\pi a}$ into asymptotic forms $f_x(k)\simeq
K(\sqrt{1-k^2})-1$, $f_z(k)\simeq 2/k^2$ with $k\ll 1$ in
Eq.~(12) gives the following from Eq.~(11):
\begin{equation}
\gamma\simeq 2H^2_0ab\left(\frac{\pi}{2}-1+\frac{1}{2}\ln\frac{4\pi
a}{c}\right).
\end{equation}

The lower inset in Fig.~3 represents the comparison of
$\gamma$ values obtained from the general formula (11) and
the asymptotic one (14).

Let us now consider the influence of a finite crystal length ($b$
dimension) on the $\gamma$ quantity. Taking an arising additional
contribution from $ac$-faces of the crystal (Fig.~1b) into account and
assuming a tangential field component on these faces to be the same as
that on the other lateral $bc$-faces, from Eq.~(11) we obtain:
\begin{equation}
\gamma=2H^2_0(abf_x+bcf_z+acf_z).
\end{equation}

However, the estimation (15) gives an overestimated value
of $\gamma$ in the case of a real three-dimensional sample.
Indeed, the limitation of the crystal length $b$ will
result in a decrease of the magnetic field tangential
component ${\bf H_t}\parallel a$ on the sample surface. The
appropriate decrease in $\gamma$ will not be compensated by
the appearance of ${\bf H_t}\parallel b$ component, which
is absent when $b\rightarrow\infty$. Thus, formula (15) is
an upper limit of $\gamma$, and its $c/a$ dependence is
shown in the lower inset in Fig.~3.

In order to check the accuracy of the above calculations we
have measured both magnetic moment $M$ and geometrical
factor $\gamma$ of superconducting slabs with different
$a/c$ and $b/a$ ratios using $ac$-susceptibility and cavity
perturbation techniques. The discrepancy between
theoretical and experimentally obtained values of $M$
reached up to $20\%$ in case of short $(b\sim 3a)$ samples
and decreased to less than $5\%$ for substantially long
samples $(b\sim 6a)$. In contrast to magnetic moment, the
values of $\gamma$ proved to be in better agreement with
the theory. In fact, divergence have never exceeded $5\%$,
probably, due to weaker (logarithmic) dependence of
$\gamma$ on $a/c$ ratio.

Both $M$ and $\gamma$ integrals in Eqs.~(8) and (9) are
convergent. At the same time, as follows from Eqs.~(5) and
(6), the magnetic field diverges $\propto 1/r^\frac 1 3$ at
small distances $r$ from the edges (point 3 in the Fig.~2).
In this connection the question arises about possible
nonlinearities on the edges of the superconducting slab. In
the resonant circuits we applied small amplitude
($H_0<0.1$~Oe) of the $ac$ magnetic field. The one-order
increase of $H_0$ did not give rise to any nonlinear
effects under the measurements of YBa$_2$Cu$_3$O$_{6.95}$
crystals.

\section{Electrodynamics of anisotropic crystal}

The electrodynamics of a layered anisotropic HTSC is characterized by
components $\sigma_{ab}$ and $\sigma_c$ of the conductivity tensor. In the
normal state ac field penetrates along the $c$-axis through a skin depth
$\delta_{ab}=\sqrt{2/\omega\mu_0\sigma_{ab}}$ and in the CuO$_2$ plane
through $\delta_c=\sqrt{2/\omega\mu_0\sigma_c}$. In the superconducting
state all parameters $\delta_{ab}$, $\delta_c$,
$\sigma_{ab}=\sigma'_{ab}-i\sigma''_{ab}$, and $\sigma_c=\sigma'_c
-i\sigma''_c$ are complex. At $T<T_c$, if $\sigma'\ll \sigma''$ the field
penetration depths are given by formulas
$\lambda_{ab}=\sqrt{1/\omega\mu_0\sigma''_{ab}}$,
$\lambda_c=\sqrt{1/\omega\mu_0\sigma''_c}$. In the close neighborhood of
$T_c$ decay of the magnetic field in a superconductor is characterized by
functions $\rm {Re}\,(\delta_{ab})$ and $\rm {Re}\,(\delta_c)$, which turn
to $\delta_{ab}$ and $\delta_c$ at $T\ge T_c$, respectively.

In the $T$-orientation the surface impedance $Z_{ab}$ is
directly connected with the in-plane penetration depth
$\lambda_{ab}(T)$ at $T<T_c$ and the skin depth
$\delta_{ab}(T)$ at $T\ge T_c$. Both lengths are smaller
than the typical crystal thickness. Hence, when the crystal
is in the $T$-orientation and at an arbitrary temperature
the surface impedance $Z_{ab}$ is defined as a coefficient
in Leontovich boundary condition\cite{Lan}, and is
correlated with the conductivity $\sigma_{ab}$ through the
local relation:
\begin{equation}
Z_{ab}=R_{ab}+iX_{ab}=\left(\frac{i\omega \mu_0}{\sigma_{ab}}\right)^{1/2}
\end{equation}

In case the HTSC microwave conductivity is real in the normal state the
real and imaginary parts of the surface impedance are equal. Hence, in the
$T$-orientation the constant $f_0$, essential to determine $X_{ab}(T)$ in
Eq.~(1), may be found as a result of $R_{ab}(T)$ and $\Delta X_{ab}(T)$
coincidence at $T\ge T_c$. It should be pointed out that thermal expansion
of the crystal may essentially affect the $X_{ab}(T)$ shape of $X_{ab}(T)$
curve in the $T$-orientation. Since the resonance frequency depends on the
volume occupied by the field, the crystal expansion is equivalent to a
reduction in the magnetic field penetration depth and results in an
additional frequency shift $\Delta f_l(T)$ of the cavity \cite{Tru2}:
\begin{eqnarray}
\Delta f_l(T)=\frac{f\mu_0}{8W} \int_s \Delta
l_i(T)H_t^2ds=\nonumber\\=\frac{f\mu_0vH_0^2}{4W}\left[\varepsilon_c
f_x +(\varepsilon_a + \varepsilon_b)f_z\right],
\end{eqnarray}
where $\varepsilon_i$ is a relative change $\Delta l_i/l_i$ of the
sample dimension $l_i$ $(a,b,c)$ resulting from the thermal
expansion, and the functions $f_x$ and $f_z$ are defined according
to Eq.~(12). In Ref.~\onlinecite{Tru2} the contribution (17) to
the overall frequency shift is shown to be negligible at low
temperatures, however, it becomes noticeable at $T>0.9\,T_c$ in
the $T$-orientation.

In the $L$-orientation at $T<0.9\,T_c$ the penetration depth in an HTSC
crystal is still smaller than characteristic sample dimensions. It allows
to treat the experimental data in terms of impedance $Z_{ab+c}$ averaged
over the sample surface in accordance with Eq.~(4). In particular, taking
account of the measurements of $\Delta\lambda_{ab}(T)=\Delta
X_{ab}(T)/\omega\mu_0$ in the $T$-orientation and the measured value
$\Delta\lambda_{ab+c}(T)=\Delta X_{ab+c}(T)/\omega\mu_0$ in the
$L$-orientation, we obtain
\begin{equation}
\Delta\lambda_c=\left[(a+c)\,\Delta\lambda_{ab+c}-
a\,\Delta\lambda_{ab}\right]/c~. \label{DL}
\end{equation}

This technique of $\Delta\lambda_c(T)$ determination was
used in microwave experiments
\cite{Kit1,Jac1,Hos,Shib1,Mao,Bon1,Shib2,Srik} at low
temperatures $T<T_c$. Nevertheless, this approach to
investigation of the surface impedance anisotropy in HTSC
crystals at $T<T_c$ does not allow to determine the value
of $\lambda_c(T)$ from the measurements of quality factor
and resonance frequency shift, nor may it be extended to
the range of higher temperatures. The point is that the
size effect provides with an essential influence in the
$L$-orientation at $T>0.9\,T_c$, when the penetration
depths $\lambda_c$ and $\delta_c$ turn out to be comparable
with the crystal width. As a result, the $R_{ab+c}(T)$
temperature dependence measured in the normal state does
not coincide with $X_{ab+c}(T)$, which makes the previous
method of determining $f_0$ non-applicable.

In this case in order to analyze our measurements in both
the superconducting and normal states we shall use the
formulae for field distribution in an anisotropic long
strip ($b\gg{a,c}$) in the $L$-orientation \cite{Gou}.
These formulae neglect the effect of the $bc$-faces of the
crystal (Fig.~1a), but allow for the size effect correctly.
At an arbitrary temperature the measured quantities
$\Delta(1/Q)$ and $\Delta f(T)=\delta f(T)+f_0$ are
expressed in terms of a complex function
$\mu(T)=\mu'(T)-i\mu''(T)$ \cite{Shov,Tru4}:
\begin{equation}
\Delta\left(\frac 1 Q\right)-2i\frac{\delta
f}f=\frac{i\mu_0\mu vH_0^2}{2W},
\end{equation}
which is controlled by the components $\sigma_{ab}(T)$ and $\sigma_c(T)$
of the conductivity tensor through the complex penetration depths
$\delta_{ab}$ and $\delta_c$:
\begin{eqnarray}
\mu={8 \over \pi^2}\sum_n {1\over n^2}\left\{
{\tan(\alpha_n) \over \alpha_n}+ {\tan(\beta_n) \over
\beta_n} \right\}&,& \nonumber \\ \alpha_n^2=-{a^2 \over
\delta_c^2} \left({i\over 2}+{\pi^2 \over 4}{\delta_{ab}^2
\over c^2}n^2 \right)&,& \nonumber \\ \beta_n^2=-{c^2 \over
\delta_{ab}^2} \left({i\over 2}+{\pi^2 \over 4}{\delta_c^2
\over a^2}n^2 \right)&,&
\end{eqnarray}
where the sum is performed over odd integers $n>0$.

If $\sigma'\ll \sigma''$ in the superconducting state, we get
\begin{eqnarray}
\mu'\simeq{8 \over \pi^2}\sum_n {1\over n^2}\left\{
{\tanh(\tilde\alpha_n) \over \tilde\alpha_n}+
{\tanh(\tilde\beta_n) \over \tilde\beta_n} \right\}&,&
\nonumber \\ \label{eq3} \tilde\alpha_n^2={a^2 \over
\lambda_c^2} \left({1 \over 4}+{\pi^2 \over
4}{\lambda_{ab}^2 \over c^2}n^2 \right)&,&  \nonumber \\
\tilde\beta_n^2={c^2 \over \lambda_{ab}^2} \left({1 \over
4}+{\pi^2 \over 4}{\lambda_c^2 \over a^2}n^2 \right)&.&
\end{eqnarray}

In particular, $\lambda_{ab}\ll c$ and $\lambda_c\ll a$ at $T<0.9\,T_c$,
so we derive from Eq.~(21) a simple expression for the real part of $\mu$:
\begin{equation}
\mu'=2\lambda_c/a+ 2\lambda_{ab}/c. \label{CH}
\end{equation}

One can easily check up that in the range of low
temperatures the change in $\Delta\lambda_c(T)$ prescribed
by Eq.~(22) is identical to that in Eq.~(18).

In the normal state the conductivity $\sigma'\gg \sigma''$.
If the sample dimensions were much more than the
penetration depths, we would obtain the following from
Eqs.~(19), (20):
\begin{equation}
\Delta\left(\frac 1 Q\right)-2i\frac{\delta f}f\simeq
(1+i)\frac{\mu_0vH_0^2}{2W}\left({\delta_c\over
a}+{\delta_{ab}\over c}\right),
\end{equation}
i.e. the temperature dependences of $\Delta(1/Q)$ and
$(-2\delta f/f)$ would be identical at $T>T_c$, and the
values $R_{ab+c}(T)$ and $X_{ab+c}(T)$, derived from
Eqs.~(2)-(4), would be equal. In practice, the value
$\delta_c\sim 0.1$~mm ($f\sim 10$~GHz) proves to be
comparable with the crystal width $a$ even for
YBa$_2$Cu$_3$O$_{6.95}$ crystals, which can be referred to
as weakly anisotropic in comparison with other layered
HTSC's. So to determine the surface impedance components
from the measurements in the $L$-orientation it is
necessary to use the general formulae (19), (20), as shown
below.

\section{Experimental results}

YBa$_2$Cu$_3$O$_{6.95}$ single crystals were grown using the
method of slow cooling from a solution-melt with the use of a
BaZrO$_3$ crucible. The initial mixture was prepared from a
mixture of oxides with mass portions Y$_2$O$_3$ : BaO$_2$ : CuO =
1 : 25 : 24 and subsequent pressing of the compound into a tablet
of 40 mm in diameter under the pressure of 200 MPa. The initial
components purities were 99.95\% for both yttrium and copper
oxides and 99.90\% for barium peroxide. Crucible material porosity
(2\%) was taken into account when choosing a heating regime and
homogenization time. Preliminary experiments have demonstrated
that the melt under use saturates the crucible walls through the
whole width (3~mm) during the period of 5-7 hours at the working
temperature. In 10 hours crystals growth terminates due to a
complete vanishing of the melt from the crucible. To reduce the
melt homogenization time which amounts to 10-20 hours at
1030$^\circ$C, according to Ref.~\onlinecite{Erb}, the method of
accelerated-decelerated rotation of the crucible \cite{Shulz} was
used, which made intensive mixing of the melt possible. The
homogenization time of the melt at 1010$^\circ$C did not exceed
one hour. Crystals growth time amounted to 2 hours at a cooling
rate of 3-4$^\circ$C/h, after which the remaining melt was
decanted at 950$^\circ$C and cooled down to room temperature at a
rate of 15-20$^\circ$C/h. The crystals obtained were saturated
with oxygen at 500$^\circ$C in an oxygen flow, after which their
critical temperature was equal to 92~K. The measurements of the
dynamic susceptibility showed that the width of the
superconducting transition in the samples did not exceed 0.1~K at
100~KHz.

The surface impedance was measured using the hot-finger
technique \cite{Tru2} at a frequency of $f=9.42$~GHz in the
$T$- and $L$-orientations.

Fig.~4 displays the typical temperature dependences  of
$R_{ab}(T)$ and $X_{ab}(T)$ for YBa$_2$Cu$_3$O$_{6.95}$
single crystal in the normal and superconducting states
measured in the $T$-orientation. The sample represented a
prolate parallelepiped with dimensions $a\times b\times
c=0.4\times 1.6\times 0.1$~mm$^3$. The sample geometrical
factor $\Gamma =90$~k$\Omega$ was calculated from Eqs.~(11)
and (3).
%fig.4-----------------------------------------
\begin{figure}[tp]
\includegraphics[width=0.45\textwidth,clip]{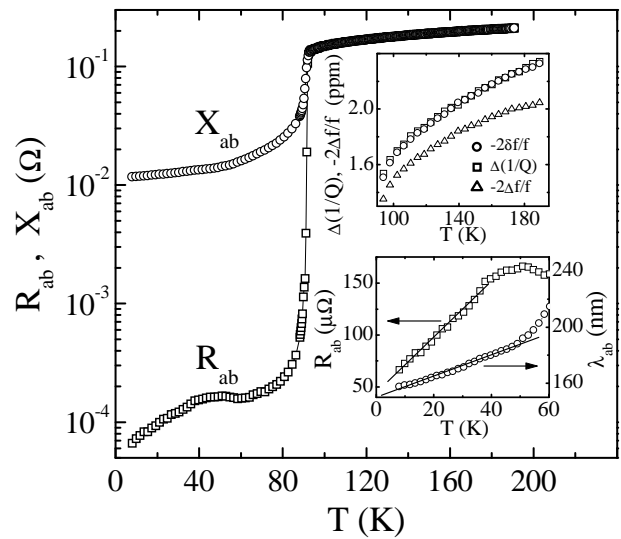}
%\vspace{0.2cm}
\caption{$R_{ab}(T)$ and $X_{ab}(T)$ of YBa$_2$Cu$_3$O$_{6.95}$
single crystal ($T$-orientation). The upper inset shows the
measured temperature dependences $-2\Delta f/f$ (triangles) and
$\Delta (1/Q)$ (squares). Taking the constant $f_0$ and thermal
expansion into account we get $-2\delta f/f$ (circles). The lower
inset displays $R_{ab}(T)$ and $\lambda_{ab}(T)$ dependences at
low $T$.} \label{f4}
\end{figure}

The upper inset in Fig.~4 displays the measured temperature
dependences of $\Delta Q^{-1}$ (squares) and $-2\Delta f/f$
(triangles) in the normal state of the crystal. The curves
$\Delta Q^{-1}(T)$ and $-2\delta f(T)/f=-2[\Delta
f(T)+\Delta f_l(T)+f_0]/f$ (circles) coincide after taking
into account the additional frequency shift $\Delta f_l(T)$
from Eq.~(17), which arises due to the sample thermal
expansion \cite{Tru2,Mein2}, along with the constant $f_0$,
which is independent of temperature. The coincidence of
$\Delta Q^{-1}(T)$ and $-2\delta f(T)/f$ curves and
equality $R_{ab}(T)=X_{ab}(T)$ at $T>T_c$ (according to
Eq.~(2)) demonstrate the fulfilling of the normal skin
effect condition in the $ab$-planes of
YBa$_2$Cu$_3$O$_{6.95}$ crystal in the $T$-orientation. A
linear temperature dependence of resistivity
$\rho_{ab}(T)=1/\sigma_{ab}(T)=0.63\,T$~$\mu\Omega\cdot$cm
in the range $100\leq T<200$~K together with the skin depth
$\delta_{ab}(150$~K)=5~$\mu$m are derived from Eq.~(16).

The temperature dependence $R_{ab}(T)$ has a broad peak in
the range $T\sim T_c/2$, characteristic of
YBa$_2$Cu$_3$O$_{6.95}$ crystals in the superconducting
state. The dependences $R_{ab}(T)$ and
$\lambda_{ab}(T)=X_{ab}(T)/\omega\mu_0$ are linear at
$T<T_c/3$ (lower inset in Fig.~4). Their extrapolation to
$T=0$~K results in the values of residual surface
resistance $R_{ab}(0)\approx 40$~$\mu\Omega$ and
penetration depth $\lambda_{ab}(0)\approx 150$~nm.

Equations (19) and (20) are used to analyze the
experimental data in the $L$-orientation of the crystal. In
the normal state the real part of Eq.~(19) defines the
relationship between $\Delta Q^{-1}(T)$ and the skin depths
$\delta_{ab}(T)$ and $\delta_c(T)$.
%fig.5-----------------------------------------
\begin{figure}[bp]
\includegraphics[width=0.45\textwidth,clip]{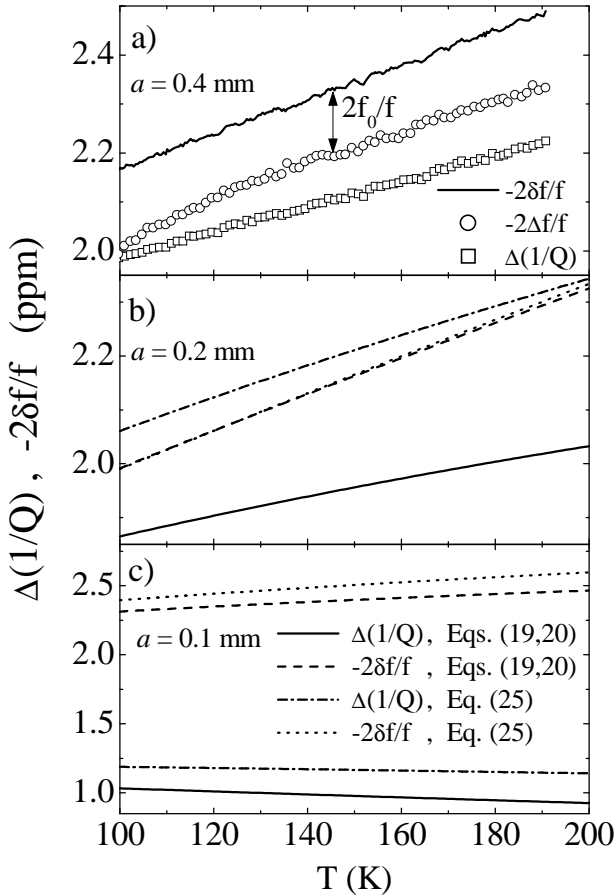}
%\vspace{0.6cm}
\caption{a) $\Delta Q^{-1}$ (squares) and $-2\Delta
f/f$ (circles) measured in the $L$-orientation of
YBa$_2$Cu$_3$O$_{6.95}$ single crystal at $T\geq 100$~K. Solid
line shows the temperature dependence of $-2\delta f/f$ derived
from Eqs.~(19), (20). The constant $2f_0/f$ is indicated by arrow.
b) $\Delta Q^{-1}$ and $-2\delta f/f$ calculated from Eqs.~(19),
(20) (solid and dashed lines, respectively) and from Eq.~(25)
(dash-dotted and dotted lines) for a sample width $a=0.2$~mm. c)
The same as b) but for $a=0.1$~mm.} \label{f5}
\end{figure}
Upon measuring the dependence $\Delta Q^{-1}(T)$ (squares in
Fig.~5a) at $T>T_c$ in the $L$-orientation and taking the
dependence $\delta_{ab}(T)=\sqrt{2\rho_{ab}(T)/\omega\mu_0}$
measured in the $T$-orientation into account, from Eq.~(20) we
obtain the function $\delta_c(T)=\sqrt{2\rho_c (T)/\omega\mu_0}$,
$\delta_c (150$~K)=0.06~mm, and the temperature dependence $\rho_c
(T)=10^4+24\,T$~$\mu\Omega\cdot$cm in the range $100\leq T<200$~K.
Using already determined dependences $\delta_c(T)$ and
$\delta_{ab}(T)$ and having computed the real part $\mu'(T)$ in
Eq.~(20), from Eq.~(19) we calculate $(-2\delta f/f)$ versus
temperature at $T>T_c$, which is shown by solid line in Fig.~5a.
This line is approximately parallel to the experimental curve
$-2\Delta f(T)/f$ in the $L$-orientation (triangles in Fig.~5a) at
$T>110$~K. The difference $-2(\delta f-\Delta f)/f$ yields the
additive constant $f_0$. Given $f_0$ and $\Delta f(T)$ we obtain
$\delta f(T)$ in the entire temperature range in the
$L$-orientation.

It should be emphasized that the discrepancy of the curves $\Delta
Q^{-1}(T)$ and $-2\delta f(T)/f$ at $T>T_c$ in the $L$-orientation does
not originate from the thermal expansion of the crystal essential in the
$T$-orientation, but arises due to the size effect. This discrepancy
becomes more noticeable with the decrease in the crystal width $a$, when
it becomes equal to the skin depth $\delta_c$. The computational result
for $\Delta Q^{-1}(T)$ (solid line) and $-2\delta f(T)/f$ (dashed line)
from Eqs.~(19) and (20) for the above mentioned dependences $\rho_{ab}(T)$
and $\rho_c (T)$ is shown in Fig.~5b and 5c for $a=0.2$~mm and $a=0.1$~mm.

It should be noticed that the surface impedance components
$R_{ab+c}$ and $X_{ab+c}$ in Eq.~(4) cannot be found from
the values of  $\Delta (1/Q)$ and $\delta f/f$ measured at
$T>0.9\,T_c$ in the $L$-orientation with the use of Eq.~(2)
due to the size effect in anisotropic HTSC crystals.
Moreover, it is also incorrect to substitute the values $R$
and $X$ in these formulae by their effective values
$R^{eff}(d)$ and $X^{eff}(d)$ for a thin metal plate of
width $d\sim\delta$, placed in microwave field ${\bf
{H}}_\omega$ parallel to its infinite surfaces
\begin{eqnarray}
R^{eff}(d)=R\,\frac{\sinh\eta-\sin\eta}{\cosh\eta+\cos\eta},&&\nonumber\\
X^{eff} (d)
=R\,\frac{\sinh\eta+\sin\eta}{\cosh\eta+\cos\eta},&&
\end{eqnarray}
where $\eta=\omega\mu_0d/2R$, $R=\sqrt{\omega\mu_0\rho/2}$. The point is
that though allowing the usage of formulae (24) in Eqs.~(2) and (4), the
solution of Maxwell equations results in an incorrect (one-dimensional)
distribution of high frequency currents in the crystal. Indeed, for
$b\parallel {\bf H_{\omega}}$ we obtain from Eqs.~(2)-(4):
\begin{equation}
\Delta\left(\frac 1 Q\right)-2i\frac{\delta
f}f=\frac{H_0^2}{\omega
W}\left[abZ_{ab}^{eff}(c)+bcZ_c^{eff}(a)\right],
\end{equation}
where the effective values
$Z_{ab}^{eff}(c)=R_{ab}^{eff}(c)+iX_{ab}^{eff}(c)$ and
$Z_c^{eff}(a)=R_c^{eff}(a)+iX_c^{eff}(a)$ are defined by
Eq.~(24). Figs.~(5b) and (5c) display the computational
result for $\Delta Q^{-1}(T)$ and $-2\delta f(T)/f$ from
Eqs.~(24), (25), and above determined dependences
$\rho_{ab}(T)$ and $\rho_c (T)$ in the case of two plates
of dimensions $a\times b\times c=0.2\times 1.6\times
0.1$~mm$^3$ and half the width $a=0.1$~mm. Presently,
having compared these results with the ones obtained from
Eqs.~(19), (20), we can see that at $a\approx 3\delta_c$
(Fig.~5b) the approximation (25) proves to be practically
insensitive to the size effect, giving rise to weakly
differing dependences  $\Delta Q^{-1}(T)$ and $-2\delta
f(T)/f$ at $T>100$~K. Only in the case of the crystal width
$a=0.1$~mm ($a\sim\delta_c$, Fig.~5c) does the
approximation (25) give rise to the result resembling the
one obtained with the aid of the formulae (19), (20).

Upon finding the dependences $\Delta Q^{-1}(T)$, $\delta
f(T)$ and, hence, the function $\mu (T)$ in Eq.~(19) in the
normal and superconducting states of
YBa$_2$Cu$_3$O$_{6.95}$ crystal in the $L$-orientation and
using the conductivities $\sigma'_{ab}(T)$ and
$\sigma''_{ab}(T)$ found from Eq.~(16), we get the $c$-axis
conductivity components $\sigma'_c(T)$ and $\sigma''_c(T)$
from Eq.~(20).
%fig.6-----------------------------------------
\begin{figure}[tp]
\includegraphics[width=0.45\textwidth,clip]{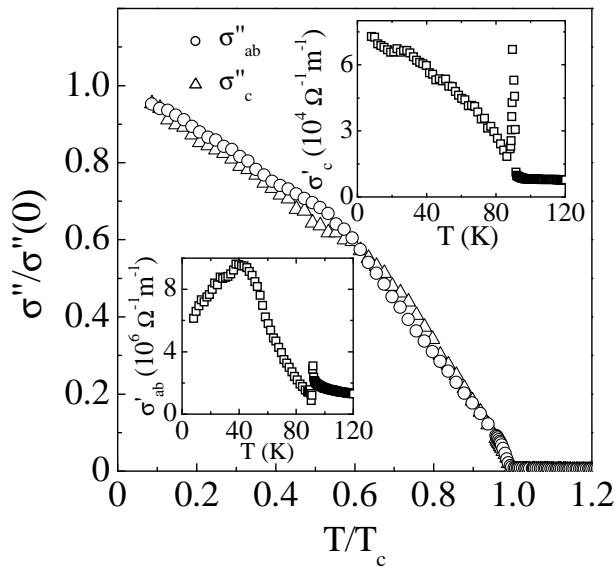}
%\vspace{0.2cm}
\caption{The conductivities $\sigma_{ab}''/\sigma_{ab}''(0)$ and
$\sigma_{c}''/\sigma_{c}''(0)$ versus reduced temperature $T/T_c$.
The insets display $\sigma_{ab}'(T)$ and $\sigma_{c}'(T)$
dependences.} \label{f6}
\end{figure}
All the conductivity tensor components
obtained are shown in Fig.~6. In case of local relationship
between the electric field and the current along the
$c$-axis, the surface impedance $Z_c(T)=R_c(T)+iX_c(T)$ is
related to the conductivity
$\sigma_c(T)=\sigma'_c(T)-i\sigma''_c(T)$ through Eq.~(1).
Fig.~7 displays the components of $Z_c(T)$ obtained in this
manner. $R_c(T)$ and $\lambda_c(T)=X_c(T)/\omega\mu_0$
dependences (insets in Fig.~7) demonstrate linear behaviour
at $T<T_c/2$. The value of the penetration depth along
cuprate planes of YBa$_2$Cu$_3$O$_{6.95}$ single crystal is
equal to $\lambda_c(0)\approx 1.55$~$\mu$m when
extrapolated to zero temperature.

From Eq.~(4) one can easily estimate that in the $L$-orientation
of our crystal the contribution of the $c$-axis currents into
measurable quantities is about two times greater than that of the
$ab$-plane ones: $bcZ_c\approx 2abZ_{ab}$. Taking the accuracy of
determination of $R_{ab}(T)$ ($<5\%$) and $\Delta\lambda_{ab}(T)$
(a few angstroms) values into account, we conclude that linear
behaviour of $R_{c}(T)$ and $\lambda_c(T)$ at low temperatures is
the property of optimum-doped YBa$_2$Cu$_3$O$_{6.95}$ and is not
due to the inaccuracy of the method used. The linear temperature
dependence of $\lambda_c$ at $T<T_c/2$ is also confirmed by the
previous microwave\cite{Mao,Srik} and low-frequency\cite{Panagop}
measurements of YBa$_2$Cu$_3$O$_{6.95}$. In contrast to the result
of Ref.~\onlinecite{Hos} we did not observe an upturn in $R_c(T)$
at low temperatures. Our surface resistance $R_c(T)$ measurements
are in qualitative agreement with other microwave
data\cite{Kit1,Mao}.
%fig.7-----------------------------------------
\begin{figure}[hbtp]
\includegraphics[width=0.46\textwidth,clip]{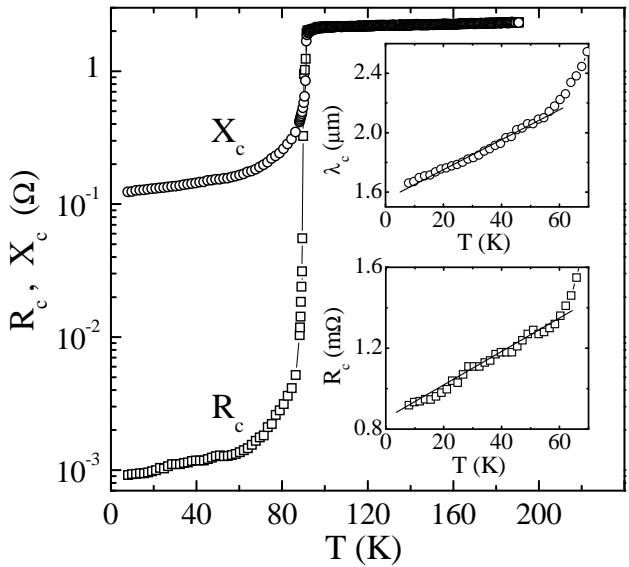}
%\vspace{0.2cm}
\caption{$R_c(T)$ and $X_c(T)$ of YBa$_2$Cu$_3$O$_{6.95}$ single
crystal. The insets demonstrate $R_{c}(T)$ and $\lambda_{c}(T)$ at
low $T$.} \label{f7}
\end{figure}

\section{Conclusion}

In conclusion, we have developed the electrodynamic
approach for HTSC surface impedance anisotropy
measurements. The major problems in analyzing these
measurements in HTSC single crystals are: (i) determining
the crystal geometrical factor in the $T$-orientation
(${\bf H_{\omega}}\perp ab$), which is dependent on the
microwave field distribution on the sample surface; (ii)
allowing for the size effect, which influences the
quantities measured in the $L$-orientation (${\bf
H_{\omega}}\parallel ab$) at $T>0.9\,T_c$.

We have shown that both these problems may be solved in the case of a
crystal in the form of a plate with dimensions $b\gg a>c$. Furthermore, in
this case we have calculated the magnetic field distribution on the
crystal surface in the $T$-orientation and found the simple expression
(11) for its geometric factor. In the $L$-orientation of the crystal, we
have shown that only the use of general formulae (19) and (20) allows to
take account of the size effect correctly and to determine the complex
conductivity and the surface impedance in the $c$-direction of the crystal
in the normal and superconducting states. The experimental technique of
measuring all the components of the conductivity tensor is described in
detail, and we hope that this technique will be useful in comprehensive
studies of anisotropic characteristics of HTSC crystals.

The reported electrodynamic approach has been successfully
applied to the analysis of the microwave ($f=9.4$~GHz)
response measurements in both the $T$- and $L$-orientations
of YBa$_2$Cu$_3$O$_{6.95}$ single crystal, grown in
BaZrO$_3$ crucible. The temperature dependences of all
conductivity and surface impedance tensors components
proved to be linear at $T<T_c/2$. Their extrapolation to
zero temperature gives the values of residual surface
resistance $R_{ab}(0)\approx 40$~$\mu\Omega$ and
$R_c(0)\approx 0.8$~m$\Omega$ and magnetic field
penetration depth $\lambda_{ab}(0)\approx 150$~nm and
$\lambda_c(0)\approx 1.55$~$\mu$m.

\section{Acknowledgments}

We thank P.~Monod for helpful discussions. This research
has been supported by RFBR grants 00-02-17053, 02-02-06578,
02-02-08004 and Government Program on Superconductivity
(contract 540-02). M.R.T. thanks Russian Science Support
Foundation.

%\begin{references}

\end{document}